\documentclass[eqsecnum,preprint,showpacs,amsmath,aps,nofootinbib]{revtex4}

\usepackage{epsfig}
\usepackage{amssymb}
\usepackage[dvips]{color}

\begin{document}
\title{Towards weighing the condensation energy to ascertain the Archimedes force of vacuum}

\author{Enrico Calloni} 
\email[E-mail: ]{enrico.calloni@na.infn.it}
\affiliation{Universit\`a di Napoli Federico II, 
Dipartimento di Fisica, Complesso Universitario di Monte S. Angelo,
Via Cintia Edificio 6, 80126 Napoli, Italy\\
Istituto Nazionale di Fisica Nucleare, Sezione di
Napoli, Complesso Universitario di Monte S. Angelo,
Via Cintia Edificio 6, 80126 Napoli, Italy}

\author{Martina De Laurentis} 
\email[E-mail: ]{martina.delaurentis@na.infn.it}
\affiliation{Universit\`a di Napoli Federico II, 
Dipartimento di Fisica, Complesso Universitario di Monte S. Angelo,
Via Cintia Edificio 6, 80126 Napoli, Italy\\
Istituto Nazionale di Fisica Nucleare, Sezione di
Napoli, Complesso Universitario di Monte S. Angelo,
Via Cintia Edificio 6, 80126 Napoli, Italy}

\author{Rosario De Rosa} 
\email[E-mail: ]{rosario.derosa@na.infn.it}
\affiliation{Universit\`a di Napoli Federico II, 
Dipartimento di Fisica, Complesso Universitario di Monte S. Angelo,
Via Cintia Edificio 6, 80126 Napoli, Italy\\
Istituto Nazionale di Fisica Nucleare, Sezione di
Napoli, Complesso Universitario di Monte S. Angelo,
Via Cintia Edificio 6, 80126 Napoli, Italy}

\author{Luciano Di Fiore} 
\email[E-mail: ]{rosario.derosa@na.infn.it}
\affiliation{Istituto Nazionale di Fisica Nucleare, Sezione di
Napoli, Complesso Universitario di Monte S. Angelo,
Via Cintia Edificio 6, 80126 Napoli, Italy}

\author{Giampiero Esposito} 
\email[E-mail: ]{giampiero.esposito@na.infn.it}
\affiliation{Istituto Nazionale di Fisica Nucleare, Sezione di
Napoli, Complesso Universitario di Monte S. Angelo,
Via Cintia Edificio 6, 80126 Napoli, Italy}

\author{Fabio Garufi} 
\email[E-mail:]{fabio.garufi@na.infn.it}
\affiliation{Universit\`a di Napoli Federico II, 
Dipartimento di Fisica, Complesso Universitario di Monte S. Angelo,
Via Cintia Edificio 6, 80126 Napoli, Italy\\
Istituto Nazionale di Fisica Nucleare, Sezione di
Napoli, Complesso Universitario di Monte S. Angelo,
Via Cintia Edificio 6, 80126 Napoli, Italy}

\author{Luigi Rosa} 
\email[E-mail: ]{luigi.rosa@na.infn.it}
\affiliation{Universit\`a di Napoli Federico II, 
Dipartimento di Fisica, Complesso Universitario di Monte S. Angelo,
Via Cintia Edificio 6, 80126 Napoli, Italy\\
Istituto Nazionale di Fisica Nucleare, Sezione di
Napoli, Complesso Universitario di Monte S. Angelo,
Via Cintia Edificio 6, 80126 Napoli, Italy}

\author{Carlo Rovelli} 
\email[E-mail: ]{rovelli@cpt.univ-mrs.fr}
\affiliation{Aix Marseille Universit\'e CNRS, CPT, UMR 7332, 13288
 Marseille, France \\
Universit\`e de Toulon, CNRS, CPT, UMR 7332, 83957 La Garde, France}

\author{Paolo Ruggi} 
\email[E-mail: ]{ruggi@ego-gw.it}
\affiliation{European Gravitational Observatory (EGO), 
I-56021 Cascina (Pi), Italy}

\author{Francesco Tafuri} 
\email[E-mail: ]{francesco.tafuri@na.infn.it}
\affiliation{Dipartimento Ingegneria dell'Informazione, 
Seconda Universit\`a di Napoli, I-81031 Aversa (CE), Italy}

\date{\today}

\begin{abstract}

The force exerted by the gravitational field on a Casimir 
cavity in terms of Archimedes' force of vacuum is discussed, the force 
that can be tested against observation is identified and it is shown that the present 
technology makes it possible to perform the first experimental tests.  
The use of suitable high-$T_{c}$ superconductors as modulators of 
Archimedes' force is motivated. The possibility is analyzed of using gravitational wave 
interferometers as detectors of the force, transported through an optical 
spring from the Archimedes vacuum force apparatus to the gravitational 
interferometers test masses to maintain the two systems well separated. 
The use of balances to actuate and detect the force is also analyzed, the
different solutions are compared and the most important experimental issues are discussed.
    
\end{abstract}


\maketitle

\section{Introduction}
\label{intro}

One of the striking and longstanding problems of 
fundamental physics is the 
irreconcilability among the two main theories of last century, General 
Relativity and Quantum Theory. A manifestation of this tension is the value 
that quantum field theory attributes to the vacuum energy density, enormously 
larger than the value constrained from General Relativity by considering the 
radius of our universe.
This problem, known as the cosmological constant problem
\cite{weinberg_art}, has been faced 
over the last decades with profound theoretical investigations, following also 
the evolution of the most important quantum gravity theories, like string 
theories, loop quantum gravity and many others 
\cite{Rovelli,Esposito,Kiefer}. None of the theoretical efforts has so far 
succeeded in finding a consensual solution, so that it is 
still questionable whether vacuum energy does interact with 
gravity, and what is its contribution to the cosmological constant 
\cite{rovelli2,why_not?}.
In spite of the common belief by the scientific
community in the existence of an interaction between vacuum energy
and gravity, not a single experimental test of this interaction exists. 

About a decade ago, it was pointed out that a possible way 
to verify the interaction of vacuum fluctuations with gravity was to 
weigh a (suitably realized, layered) rigid Casimir cavity 
\cite{vacuum_fluctuation_force}. At that time it was yet unclear whether 
Casimir energy could be modulated in a rigid cavity. Furthermore, the most 
important macroscopic  detectors of exceedingly small forces, the 
gravitational wave detectors with which we compared our force, were still 
under construction. Nowadays, 
thanks to many activities in the various fields mentioned, 
the situation has been remarkably improved 
so that it is possible to step from the initial 
idealistic experiment to a road map towards the measurement of the effect.

The paper is organized as follows. In Sec. II  the theory of the experiment 
is recalled and discussed to clearly identify the measured quantity. In 
Sec. III, the need for superconductors as actuators of  
modulation of the Casimir stress-energy tensor is proposed.
In particular, the use of High-$T_{c}$
superconductors is pointed out. The first part of the section is devoted to describe the theory and method of the evaluation of vacuum 
energy in the well-established case of type-I superconductors. This makes it possible to discuss, in the second part, the hypothesis 
and approximations  assumed for the case of type-II superconductors.  
Finally, the force exerted by gravity on a multi-layer superconductor Casimir-cavity system will be considered.  
In Sec. IV the expected force is compared with the sensitivity of advanced gravitational wave detectors: an optical technique to link 
the force on the Casimir-test mass with the test mass of  gravitational wave detectors is presented and the noises 
are discussed in details. In Sec. V the possibility to perform the 
measurement in the superconductors' transition-favored 
low-frequency regime is discussed by 
analyzing the use of a suitable seismic isolated balance. The comparison 
of the two experimental ways is discussed in light of the most critical 
experimental issues. 

\section{Theoretical aspects}
\label{Theory}

Let us consider a rigid Casimir cavity in a weak gravitational field, like 
the one, for instance, of a laboratory at rest on the surface of the earth. 
To first order, the reference system is the Fermi system for which, 
neglecting rotations, one can write the line element  
as \cite{vacuum_fluctuation_force,USA1}: 
\begin{equation}
ds^{2}=-(1+2A_{j}x^{j})(dx^{0})^{2}+\delta_{jk}dx^{j}dx^{k}
+O_{\alpha \beta}(|x^{j}|^{2})dx^{\alpha}dx^{\beta}.
\end{equation}
The term $c^{2}{\vec A}$, $c$ being the speed of light, is the observer's acceleration with respect to the 
local freely falling frame. It has components $(0,0,|{\vec g}|)$), where 
$g$ is the gravitational acceleration.   The term 
$-2A_{j}x^{j}$ is proportional to distance along the acceleration direction; 
$x^{3}$, which we also denote by $z$, is positive in the upwards 
direction.

The force exerted by the gravitational field on a rigid Casimir cavity, 
with plates of proper area $\cal A$, separated by the proper distance $a$ 
and placed orthogonal to the gravitational acceleration $\vec g$, has been 
calculated in different ways \cite{vacuum_fluctuation_force,
energy-momentum_tensor,USA1,USA2,USA3,USA5,relativistic_mechanics}. To clarify the proposed measurement we briefly recall the main points. 

When a force is applied to a stressed body it is in general expected that also the spatial components of the stress-energy tensor contribute 
to the mass. This was firstly shown by Einstein \cite{einstein} and it is reported, for example,       
in Ref. \cite{MTW}, Eqs. (5.53), (5.54) where one considers a stressed body in a locally inertial frame, that is accelerated with 
$a^k = \frac{dv^k}{dt}$. The mass density is described by the tensor $m^{ik} = T^{\hat{0}\hat{0}}\delta^{ik} +  T^{\hat{i}\hat{k}}$, 
where the hat over the indices denotes the stress tensor in the rest frame of the medium and the force is defined as 
$F^j = \sum_{k} m^{jk}\frac{dv^k}{dt}$. 
In analogy, since the Casimir cavity is a stressed body, one could expect that the measurement of its weight would end 
in measuring the stressed-body mass and not simply the mass associated to the $T^{\hat{0}\hat{0}}$ term.
In this case, considering that the rest frame stress-energy tensor of a Casimir cavity is given by Ref. \cite{brown}:
\begin{equation}
\langle T^{\mu \nu} \rangle = \frac{\pi^{2}{\hbar}c}{ 180 a^{4}}
\left(\frac{1}{4}\eta^{\mu \nu}-{\hat h}^{\mu}{\hat h}^{\nu}
\right),
\end{equation}
where ${\hat h}^{\mu}=(0,0,0,1)$ is the unit spacelike $4$-vector
orthogonal to the plates' surface, one could expect that the cavity, of volume $V = a\cal A$, placed with plates parallel to the earth 
surface, would have a mass $m_h = V\left( T^{\hat{0}\hat{0}} + T^{\hat{3}\hat{3}} \right) = 4 a{\cal A}T^{\hat{0}\hat{0}}  
= 4\frac{E_{\rm{cas}}}{c^2}$, where $E_{\rm{cas}} = -{\cal A}\frac{\hbar \pi^2 c}{720 a^3}$ is the energy of the system, the Casimir energy. 
This would result in the force $\vec F_h = 4\frac{E_{\rm{cas}}}{c^2}\vec g$ exerted by the gravitational field on the cavity. 
Although this result is compliant with General Relativity it is nevertheless somewhat surprising, because usually one is accustomed to attributing, 
to a body of rest energy $E$, the weight $\vec F = \frac{E}{c^2} \vec g $. The surprise is indeed correct, because in fact the previous 
result does not correspond to the actual force measured in a weight experiment where the cavity is rigid and hanged at a single fixed point 
(or placed on a plate of a balance).
To correctly evaluate the force measured in these cases, the force densities acting on the various points must be red-shifted:  
Refs. \cite{USA1} and in particular \cite{relativistic_mechanics} clarified that, if the total force acting on an extended body is defined as the sum 
of red-shifted force densities, the mass  is independent of the spatial stress-energy tensor.

In order to very well clarify the measured quantity we consider the forces on each plate, expanded to first order in 
$\epsilon \equiv 2 \frac{{\rm g} a}{c^2}$, derived in Ref. \cite{energy-momentum_tensor}. 
In that work the regularized and renormalized energy-momentum tensor $T^{\mu \nu }$ has been obtained from the Hadarmard Green function of a 
Casimir apparatus in a weak gravitational field. 
The forces on the plates are the components of the resulting stress-energy tensor and, for the 
$z$-direction, have been evaluated as (hereafter $Q_2$ refers to the upper plate and $Q_1$ to the lower plate):
\begin{equation} 
\vec{f}_{Q_2}\approx\,- \frac{\pi^2}{240}
\frac{{\cal A}{\hbar} c}{a^{4}} \,\left[1-\frac{\rm g}{c^2}
\left(\frac{2}{3}\,a \right)\right]\,\hat{z}, 
\label{equazione_1}
\end{equation}
while for the lower plate we get 
\begin{equation}
\vec{f}_{Q_1}\,\approx\,\frac{\pi^2}{240} \frac{{\cal
A}{\hbar}c}{a^{4}} \left[1+\frac{\rm g}{c^2}\left(\frac{2}{3}\,a
\right)\right]\,\hat{z}. 
\label{equazione_2)}
\end{equation}

The mere addition of such forces (as it might be obtained by independently measuring the forces acting on the two plates of a 
nonrigid system) would lead to the quantity $\vec f_{{\rm ind}}$ equal to
\begin{equation}
\vec f_{{\rm ind}} = \vec{f}_{Q_{1}} + \vec{f}_{Q_{2}} \approx\, \left(\frac{\left| E_{{\rm cas}}\right| }{c^2} \, 
{\rm(g)} + F_{{\rm cas}} \delta\phi \right)\hat{z}, 
\label{nonrigid}
\end{equation}
where $P_{{\rm cas}} = {\cal A}\frac{\hbar\pi^2c }{240a^4}$ is the Casimir pressure, 
$F_{{\rm cas}} = {\cal A} P_{{\rm cas}}$ the Casimir force and where $\frac{g \, a}{c^2} = \delta\phi $ 
has been explicitly written as the variation of the gravitational potential on passing from lower to 
upper plate. By some algebra the equation (\ref{nonrigid}) reads as  $\vec f_{\rm{ind}} = 4 \frac{E_{\rm{cas}}}{c^2}\vec g$ 
corresponding to the case of nonrigid cavity. 
Interestingly, Eq. (\ref{nonrigid}) is the sum of two contributions: the 
vacuum weight part $ \frac{E_{{\rm cas}}}{c^2}{\rm g}$ 
and the Casimir pressure difference, multiplied by the surface,  
$ {\cal A} P_{{\rm cas}} \delta\phi $ on passing from one plate to the other. 
This difference in pressure is physical, and it implies the red-shifting of  
vacuum density in the gravitational field. It is similar to the 
Tolman-Ehrenfest effect \cite{tolman-ehrenfest,rovelli1} where the same 
dependence is found in the temperature of a gas at equilibrium in a 
gravitational field.

In the measurement we are interested in, however, the plates are weighed by acting on one and the same point, i.e., the suspension point 
of the rigid Casimir apparatus. In this case, as shown in Ref. \cite{relativistic_mechanics}, the gravitional red-shift 
must be taken into account when summing the force to obtain the total force acting on the body. 
By red-shifting the force up to the 
common point $Q_2$, the total force is given by (recall that $\hat{z}$ and $\vec g$ have opposite direction)
\begin{eqnarray}
\vec{F} = \vec{f}_{Q_2}+r_{Q_2}(Q_1)\vec{f}_{Q_1}^{\rm(C)}
& \approx &
F_{{\rm cas}}\left\{-\left[1-\frac{\rm g}{c^2}\left(\frac{2}{3}\,a
\right)\right]+
\left[1-\frac{\rm g}{c^2} a \right]
\left[1+\frac{\rm g}{c^2}\left(\frac{2}{3}\,a
\right)\right]\right\}\,\hat{z}
\nonumber \\
& \approx & \frac{1}{3}\frac{\rm g \, a }{c^2}\,F_{{\rm cas}}\;\hat{z}
=\frac{E_{{\rm cas}}}{c^2}\,\vec{{\rm g}}.
\label{equazione_3}
\end{eqnarray}
This condition is the case of the experiment here proposed, where a  
rigid (multi)cavity system is suspended in the gravitational field of the 
earth. This is the force that must be tested against observation and it is in full agreement with the expectation of the equivalence 
principle. It is directed upwards and it is equal to the weight of the modes of the vacuum
that are removed from the cavity. Therefore it can be interpreted as an
Archimedes buoyancy force in vacuum.

\section{Superconductors}
\label{analisysI}
The measurement of the effect cannot be performed statically. This would 
make it necessary to compare the weight of the assembled cavity with the
sum of the weights of its individual parts, which cannot be performed.
Thus, it becomes necessary to modulate the Casimir 
energy contained in the cavity to be weighed, so as to perform the 
measurement in a region of frequency where the macroscopic detectors of 
small forces have good sensitivity. 
Furthermore, to actually perform the measurement, the cavity should be a 
rigid body, so as to be weighed as a whole, and consisting of a 
multilayer of many cavities to enhance the effect.
A key point in modulation is that the energy supplied to the system should 
be at most of the same order of magnitude of the Casimir energy modulation, 
otherwise it will be extremely difficult to recover the Casimir contribution 
to the weight. Some recent techniques, as an example, 
even if very interesting for studying the Casimir force \cite{bordag,umar}, 
cannot be applied in our case because the efficiency is very low: only a 
few parts on a billion of the energy supplied to the system are converted in 
Casimir energy variation.

\begin{figure}
\includegraphics[width=0.8\linewidth]{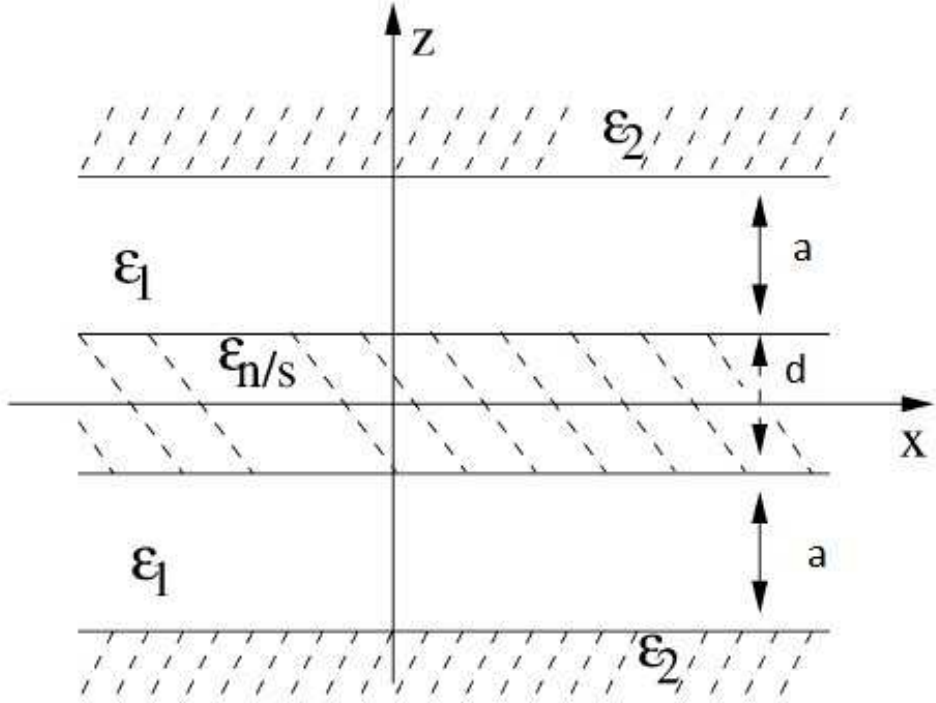}
\caption{Five-layer cavity: a thin superconducting film of thickness $d$ is placed between two thick metallic
slabs, which constitute the plates of the cavity. The gaps of width $a$ that separate the film from the plates
are filled with insulating material.}
\label{fivelayers}
\end{figure}

One possible way is to use superconductors. To show the foundation of the theory and method of evaluation of vacuum energy, in the first 
part of the section we show some known results in case of type-I superconductors. This will allow, in the second part of the section, to 
discuss both the motivation for using type-II superconductors and the present limits and approximations in evaluating the vacuum energy 
in that case. To fix the ideas consider a double cavity, consisting of
two identical plane parallel mirrors, made of a nonsuperconducting
and nonmagnetic metal, between which a plane
superconducting film of thickness D (order of few nanometers) is placed,
separated by a nonconducting material gap of equal width L (order few nanometers) from the two mirrors, as in 
Fig. \ref{fivelayers}. If the supercondutor is of type I, for any temperature $T$ lower than the transition temperature 
$T_c$ the transition Gibbs free energy $\Delta F$ can be written as the sum of the condensation energy $\mathcal{E}(T)$ and the 
variation of Casimir energy $\Delta E_{{\rm cas}}(T)$ :
\begin{equation}
\Delta F = \mathcal{E} (T) + \Delta E_{{\rm cas}}(T).
\end{equation}

In writing these equations, we have exploited the fact that
all quantities referring to the film, like the penetration
depth, condensation energy, etc., are not affected by virtual
photons in the surrounding cavity. This is a very good approximation, since the leading effect of radiative corrections
is a small renormalization of the electron mass as discussed in Refs. \cite{supertheory,variation}. 
The variation of Casimir energy at the transition can be calculated starting from the theory of Casimir energy in stratified media, 
derived in Ref. \cite{bordag}. We consider first the $T=0$ case. The Casimir energy is given by 
the sum over the cavity modes; the wave numbers $k$ are discretized 
in the $z$ direction (orthogonal to the plates) and continuous in the parallel directions (the $xy$ plane). The variation of Casimir 
energy $\Delta E_{\rm{cas}}^0(a,d)$ at the transition can then be written as
\begin{widetext} 
\begin{equation} 
\Delta E_{{\rm cas}}^0(a,d)=A\,\frac{\hbar}{2}  \int \frac{dk_1 dk_2}{(2 \pi)^2} \left\{\sum_p (\omega_{{\bf
k_\bot},\,p}^{(n,\,TM)}+\omega_{{\bf k_\bot},\,p}^{(n,\,
TE)})-\sum_p (\omega_{ {\bf k_\bot},\,p}^{(s,\,TM)}+\omega_{ {\bf
k_\bot},\,p}^{(s,\,TE)}) \right\}\;,\label{unren}
\end{equation}
\end{widetext}
where $A \gg a^2$ is the area of the cavity, ${\bf
k_\bot}=(k_1,k_2)$ denotes the two-dimensional wave vector in the
$xy$ plane, while $\omega_{{\bf k_\bot},\,p}^{(n/s,\,TM)}$
($\omega_{{\bf k_\bot},\,p}^{(n/s,\,TE)}$) denote the proper
frequencies of the TM (TE) modes,  in the $n/s$ states of the film, respectively.

By exploiting the Cauchy integral formula, and by subtracting the contribution corresponding to infinite
separation $a$ (for details, we refer the reader to chapter 4 of first item of Ref. 
\cite{bordag}),  one can rewrite the renormalized sums in Eq. (8) as integrals over complex frequencies $i\zeta$:
\begin{widetext}
\begin{equation} 
\left(\sum_p \omega_{{\bf k_\bot},\,p}^{(n,\,TM)} -\sum_p
\omega_{ {\bf k_\bot},\,p}^{(s,\,TM)}\right)_{\rm ren}=\frac{1}{2
\pi}\int_{-\infty}^{\infty} d \zeta\,  \left(\log
\frac{\Delta^{(1)}_n(i
\zeta)}{\widetilde{\Delta}^{(1)}_{n\,\infty}(i \zeta)}-\log
\frac{\Delta^{(1)}_s(i
\zeta)}{\widetilde{\Delta}^{(1)}_{s\,\infty}(i
\zeta)}\right)\;,\label{rensum}
\end{equation}
\end{widetext}
where ${\Delta}^{(1)}_{n/s}(i \zeta)$ is the expression in Eq.
(4.7) of Ref. \cite{bordag} (evaluated for $\epsilon_0=\epsilon_{n/s}$)
and $\widetilde{\Delta}^{(1)}_{n/s\,\infty}(i \zeta)$ denotes the
asymptotic value of ${\Delta}^{(1)}_{n/s}(i \zeta)$ in the limit
$a \rightarrow \infty$ (corresponding to the limit $d \rightarrow
\infty$ with the notation of Ref. \cite{bordag}). A similar expression
can be written for the $TE$ modes, which involves the quantity
${\Delta}^{(2)}_{n/s}(i \zeta)$ defined in Eq. (4.9) of
\cite{bordag}. Upon inserting Eq. (\ref{rensum}), and the
analogous expression for  $TE$ modes, into Eq. (\ref{unren}) one
gets the following expression for the (renormalized) variation
$\Delta E^{(C)}(a,d)$ of the Casimir energy:
\begin{widetext}
\begin{equation}
\Delta E_{\rm{cas}}= A \;\frac{\hbar}{2}\int   \, \frac{d {\bf k_{\bot}}}{(2 \pi)^2} \int_{-\infty}^{\infty} \frac{d
\zeta}{2 \pi}\, \,\left(\log \frac{Q_n^{TE}}{Q_s^{TE}}+\log
\frac{Q_n^{TM}}{Q_s^{TM}}\right)\;, \label{lif}
\end{equation}
\end{widetext} 
where we set 
\begin{equation} 
Q^{(TM/TE)}_I(\zeta) \equiv
\frac{\Delta^{(1/2)}_I(i
\zeta)}{\widetilde{\Delta}^{(1/2)}_{I\,\infty}(i
\zeta)}\;,\;\;\;I=n,s\,. 
\end{equation} 
The $d {\bf k_{\bot}}$ integration can be re-expressed through the $dp$ integration by means of the standard formula 
$k_\bot^2=(p^2-1)\zeta^2/c^2$. The above expression for $\Delta E_{\rm{cas}}(a,d)$ turns therefore into
\begin{widetext} 
\begin{equation} 
\Delta E_{\rm{cas}}=\frac{\hbar A}{4 \pi^2
c^2}\int_1^{\infty} p\,dp \int_0^{\infty} d \zeta\,\zeta^2 \,
\left(\log \frac{Q_n^{TE}}{Q_s^{TE}}+\log
\frac{Q_n^{TM}}{Q_s^{TM}}\right)\;, \label{denren}
\end{equation}
\end{widetext} 
where the coefficients $Q^{(TM/TE)}_I$ read as
\begin{widetext}
\begin{eqnarray}
\; & \; &
Q_I^{TE/TM}( \zeta,p) \nonumber \\
&=&
\frac{(1-\Delta_{1I}^{TE/TM}\Delta_{12}^{TE/TM}e^{-2 \zeta\,K_1  \,
L/c})^2 -(\Delta_{1I}^{TE/TM}-\Delta_{12}^{TE/TM}e^{-2 \zeta \,K_1\,
L/c})^2 e^{-2 \zeta K_I D/c}}{1-(\Delta_{1I}^{TE/TM})^2 e^{-2
\zeta K_I \,D/c}}\;,\label{delec} \nonumber \\
& \; &
\Delta_{j\,l}^{TE}=\frac{K_j- K_l}{K_j+K_l}\;,\;\;
\Delta_{j\,l}^{TM}=\frac{K_j \,\epsilon_l\,(i \zeta)-K_l
\,\epsilon_j\,(i \zeta)}{K_j\, \epsilon_l\,(i \zeta)+K_l\,
\epsilon_j\,(i \zeta)}, \nonumber \\
& K_j & =\sqrt{\epsilon_j\,(i
\zeta)-1+p^2}\;,\;\;\;I=n,s\;\;;\;\;j\,,\,l=1,2,n,s.\label{defs}
\end{eqnarray}
\end{widetext}

The generalization of these formulas to the case of finite temperature $T$ can be done with the well-known 
technique of Matsubara frequencies. This consists in replacing in Eq. (\ref{lif}) the integration 
$ \int d \zeta/2 \pi$ by the summation $k T/\hbar \sum_l$ over the
Matsubara frequencies $\zeta_l=2 \pi l/\beta$, where
$\beta=\hbar/(k T)$. This leads  to the following expression for
the variation $\Delta E_{\rm{cas}}(T)$ of  Casimir free energy:
\begin{widetext}
\begin{equation} 
\Delta E_{\rm{cas}}(T)=A\, \frac{k
\,T}{2}\sum_{l=-\infty}^{\infty}\int \frac{ d {\bf k_\bot}}{(2
\pi)^2} \,\left(\log \frac{Q_n^{TE}}{Q_s^{TE}}+\log
\frac{Q_n^{TM}}{Q_s^{TM}}\right)\;. \label{fint}
\end{equation}
\end{widetext}
Equations (\ref{denren}-\ref{fint}) involve the dielectric functions $\epsilon\,(i \zeta)$ of the various layers evaluated at 
imaginary frequencies $i \zeta$. 

For the outermost metal plates, the Drude model for the
dielectric function can be used: 
\begin{equation}
\epsilon_D(\omega_p)=1-\frac{\Omega^2}{\omega(\omega+i
\gamma)}\;,\label{drper}
\end{equation} 
where $\Omega_p$ is the plasma frequency
and $\gamma=1/\tau$, with $\tau$ the relaxation time. We denote by
$\Omega_{p2}$ and $\tau_{p2}$ the values of these quantities for the
outer plates. As is well known, the Drude model provides a very
good approximation in the low-frequency range $\omega \approx 2
k\, T_c/\hbar \simeq 10^{11}\div 10^{12}$ rad/sec which is
involved in the computation of $\Delta E_{cas}(T)$. 
The  continuation of Eq. (\ref{drper}) to the imaginary axis is of
course straightforward and gives 
\begin{equation} 
\epsilon_D(i
\zeta)=1+\frac{\Omega^2}{\zeta\,(\zeta+ \gamma)}
\;.\label{druima}
\end{equation}
For the insulating layers, a constant dielectric function can be taken, as a good approximation \cite{bordag,variation}, 
equal to the static value:
\begin{equation}
\epsilon_1(\omega)=\epsilon_1(0)\;.
\end{equation}

As far as the film is concerned, in case of type-I superconductors, 
the Drude expression, Eq. (\ref{drper}), can be used in the normal state, with appropriate values for the
plasma frequency $\Omega_n$ and the relaxation time $\tau_n$.

In  the superconducting state, the technical details are more involved, but the theory is still based on firm ground. The real part 
of the conductivity $\sigma(\omega)$ has a semi-explicit form, derived by the BCS theory, that we report in Appendix A, and shows 
the lowering of absorption component for frequencies $\hbar \omega$ less than the condensation energy gap $\Delta(T)$, and tends to the 
Drude expression for higher frequencies. (See Appendix A for details). 

From the real part of the conductivity $\sigma'(\omega)$ one can  obtain the imaginary part of the dieletric function 
$\epsilon''(\omega)$ with the standard relation  
\begin{equation} 
\epsilon''(\omega)= \frac{4
\pi}{\omega}\,\sigma'(\omega). \label{cond} 
\end{equation}
Last, from the dispersion relation, the dielectric function at imaginary frequency can be found in the form
\begin{equation} 
\epsilon_s(i \zeta)-1=\frac{2}{\pi}
\int_0^{\infty} d\omega \frac{\omega
\,\epsilon''_s(\omega)}{\zeta^2+\omega^2}\;.\label{disp}
\end{equation}.
 
With this recipe it is possible to calculate the variation of free energy at the transition.
In general, for a stand-alone superconductor, not being part of a Casimir cavity, the free energy variation at the transition is equal 
to the source magnetic energy necessary to destroy the superconductivity:
\begin{equation} 
\,\frac{V}{2\mu_0}\,\left(\frac{B_{c \|}(T)}{\rho}\right)^2 = {\cal E}_{\rm
cond}(T)\;,\label{hcri}
\end{equation}  
where V is the volume of the superconducting film. The term $\rho$ takes into account that
for a thin film, of thickness $d \ll \lambda,\xi$ (with
$\lambda$ the penetration depth and $\xi$ the correlation length),
placed in a parallel magnetic field, expulsion of the magnetic
field is incomplete, and consequently the critical field increases
from $B_c$ (the bulk value) to $B_{c \|}$. Following the Ginzburg-Landau theory, the transition is a second-order transition 
(no latent heat) and as $B$ approaches $B_{c \|}$ the order parameter (energy gap,  ''number of superconducting electrons", or 
Ginzburg-Landau $\psi$ function) approaches zero continuously while the penetration depth $\lambda$ increases from $\lambda(T)$, 
the value at zero field, to infinity \cite{tinkham}. The coefficient $\rho$ has the
approximate expression 
\begin{equation} 
\rho \approx\sqrt{24}\;
\frac{\lambda}{d}\left(1+\frac{9 d^2}{\pi^6
\xi^2}\right)\;,\label{hfilm}
\end{equation} 
where the second term inside the brackets accounts for surface nucleation.

If the film is part of a cavity, the variation of energy at the transition is the sum of the condensation energy and the Casimir energy, 
so that the previous equation becomes
\begin{equation} 
\frac{V}{2
\mu_0}\,\left(\frac{B_{c \|}^{\rm cav}(T)}{\rho}\right)^2 \,= {\cal
E}_{\rm cond}(T)\,+\,\Delta E_{\rm{cas}}(T)\;.\label{hcricav}
\end{equation} 

This equation shows that it is possible to measure the contribution of Casimir energy to the total free energy variation: it consists in 
measuring the critical magnetic field for a stand-alone film and compare it with a film that is part of a Casimir cavity. The relative shift is 
\begin{equation} 
\frac{\delta B_{c \|}}{B_{c \|}}
\approx \frac{\Delta E_{\rm{cas}}}{2\,{\cal E}_{\rm cond}(T)}.
\end{equation}
For suitable choice of the parameters, like superconductor and metal and dieletric materials, thicknesses, temperatures, it is possible 
to show experimentally that the Casimir effect enhances the critical field. The measurement has been indeed performed and shown to be  
fully compatible with the expectations \cite{low-noise}. 

The use of type-I superconductors for measuring the vacuum energy at the transition is thus meaningful and relies 
upon firm ground. Nevertheless, since the type-I superconductors are good conductors also in normal state, the modulation of Casimir energy, 
with respect to total Casimir energy, $\eta =  \frac{\Delta E_{{\rm cas}}}{\, E_{{\rm cas}}}$, is quite small, of order 
$\eta \approx 10^{-8}$ for a few nanometers thicknesses and temperatures of order 1 K \cite{variation}. With this tiny modulation 
it is possible to measure the effect on the critical field and on the variation of transition energy, because also the condensation energy, 
in type-I superconductors, is small. But it is not sufficient to prove the weight of the vacuum, because it is in absolute too small. 
It is therefore necessary to consider high-$T_{c}$ superconductors.

Some of their properties are of particular interest: generally high-Tc superconductors, 
particularly cuprates, are by construction multilayered cavities, being 
composed by Cu-O planes, that perform the superconducting transition, 
separated by nonconducting planes. More important, in normal state, 
also the Cu-O planes are poor conductors, so that the 
variation of Casimir energy is high at the transition.

In these systems the evaluation of Casimir energy is not yet completely exploited. A first important step has been the recent analysis 
on the Casimir energy of a cavity composed by two flat plasma sheets at zero temperature \cite{barton,bordag}. The theoretical foundation 
is the same as for dielectric materials and conductors described above and it is based on the summation over zero-point energies of 
electric modes. The approximations of plasma sheet, with no internal dissipation, and zero temperature give to the result the status 
of a work that can be used as providing the order of magnitude of the effect. The calculation of the renormalized energy 
$E_{{\rm cas}}$ brings thus to the usual formula for two planes separated by a distance $a$: 
\begin{equation}
E_{{\rm cas}} = -\frac{\hbar}{2c}  \int \frac{ d {\bf k_\bot}}{(2 \pi)^2} 
\int_{k_{0}}^{\infty} \frac{dk}{\pi} \frac{k}{\omega( {\bf k_\bot}, ik)} \log t(ik),
\end{equation}
where the lower integration boundary $k_{0} = {\bf k_\bot}$ and the transmission coefficients $t$, for plasma sheets, for the TE and 
TM modes are given by \cite{bordag} 
\begin{equation}
(t(ik))^{-1} = 1 - \left(\frac{\Omega}{k + \Omega}\right)^2 e^{-2ka},  \,\,\,\ (TE),
\end{equation}
and 
\begin{equation}
(t(ik))^{-1} = 1 - \left(\frac{\Omega k }{ {\bf k_\bot}^2 - k^2 - \Omega k}\right)^2 e^{-2ka},  \,\,\,\ (TM).
\end{equation}
The parameter $\Omega$ is proportional to the density of the carrier in the plasma sheet \cite{barton,bordag}:
\begin{equation}
\Omega \equiv \frac{nq^2}{2mc^2\epsilon_0} ,
\end{equation}
where $n$ is the surface density of delocalized particles, $q$ their electric charge, $m$ their mass. For small separation $a$ 
the above integrals lead to the expression for energy
\begin{equation}
E_c(a) = -5 \times 10^{-3} \hbar \frac{c A}{ a^{5/2}} \sqrt{\Omega} .
\label{ekempf} 
\end{equation}
An estimate of the parameter $\Omega$ has been proposed recently by \cite{kempf} with the aim of evaluating the Casimir effects in 
High-$T_{c}$ cuprates. The  particles' density is estimated as $n = 10^{14} \, cm^{-2}$, the charge $q = 2e$, the mass 
$m = 2\alpha m_e$ with $\alpha = 5$. Inserting these values in  Eq. (\ref{ekempf}), the reduction factor of Casimir energy with respect 
to the ideal case, at typical separation $a \approx 1$ nm turns out to be $\eta(a) = 4 \times 10^{-4} \times \sqrt{\frac{a}{1nm}}$. Considering 
that in normal state the layer is very poorly conductive, this factor is (almost) equal to the variation of Casimir energy in the transition. 
Thus, the use of High-$T_{c}$ superconductors leads to the gain of about $4$ orders of magnitude in the modulation of Casimir energy.

The other key point is the ratio between the variation in Casimir energy 
at the transition and the total energy variation.
In his paper \cite{kempf}, Kempf, checking his hypothesis with a calculation of the critical temperature $T_c$,  has conjectured that in 
cuprates the whole energy variation at the transition could be due to Casimir energy.
A check of this hypothesis can be done by comparing the estimated variation of   
Casimir energy with the total variation of the energy of the superconductor at the transition. As reported in Appendix B, in 
type-II superconductors the energy variation is determined by the thermodynamical critical field $B_c(T)$. In cuprates the critical field is 
of order of 1 Tesla (for a detailed description and calculation see Appendix B). The energy density variation $\Delta U$ is about 
\begin{equation}
\Delta U \approx  \frac{B_c^2}{2\mu_0} \approx 4 \times 10^5 \,\, J/m^3 . 
\end{equation} 
The variation of Casimir energy density $\Delta U_{\rm{cas}}$ is, following the Kempf estimate, 
\begin{equation} 
\Delta U_{{\rm cas}} \approx \eta(a) \frac{N \pi^2}{720}\frac{\hbar c}{a^{3}} \approx 2 \times 10^5 \,\, J/m^3 ,
\end{equation}
where $N \approx 10^9$ is the number of cavities per unit height. The two energies are indeed, roughly, of the same order of magnitude. 
  
Notice that, as stated in Ref. \cite{kempf}, the separation among the plates 
being of order of 1 nm, the ``Casimir" energy is dominated by plasmons 
(i.e. by the Van der Waals) energy with respect to vacuum energy. Thus, 
our assumption of Kempf's hypothesis should be regarded also as a starting 
point for further investigations on high-$T_{c}$ superconductors, to be 
performed in the near future, directed in two ways. First, regarding the present analysis as an order of magnitude estimate, 
evaluate more accurately the Casimir energy variation at the transition and its contribution to total energy; second, extend it to 
superconductors with higher spacing among conducting planes until the conditions already studied in previous measurements with 
metallic plates \cite{supertheory,variation,low-noise} are recovered.

The actual modulation of the effect can be performed in two ways: 1) by 
applying a time dependent magnetic field that spoils the 
superconducting state so as to have zero magnetization both in initial and final state; in this condition the actual measurement can be performed 
in nonvanishing applied field, because the magnetization is brought to zero and the interaction with magnetic field is minimized also at the 
final state. (A further possibility is to put the sample in two different conditions of superconductivity, with more/less regions where the 
sample is superconducting, both at vanishing applied 
field: this can be obtained by using hysteretic superconductors. Among them, 
as an example, the cuprates).
2) By temperature modulation in vanishing field. Both cases have no latent heat (see also Appendix B).

The quantity that will generate the variation of gravitational force on the sample is (the variation of) the internal energy 
$U V$, where V is the volume of the sample of the superconductor. The variation of internal energy density $U$  is evaluated for the 
two modulation cases in Appendix B. 
It is given by the equation (see \ref{dufinal})
\begin{equation}
\Delta U = \int_T^{T_c}C_ndT + \frac{B_0^2}{2\mu_0}\left[1 - (T/T_c)^2\right]^2 
+ 2 \left(\frac{T}{T_c}\right)^2 \left[1 - (T/T_c)^2\right].
\label{dufinalbis}  
\end{equation}
This is the sum of three terms: the internal energy variation of normal state (present only in case of temperature modulation), 
the contribution of the Gibbs energy and the contribution of entropy. The third term, for temperatures near $T_{c}$, gives the biggest 
contribution. This equation shows that the variation of internal energy is proportional to, and roughly of the same order of magnitude 
of the energy of the thermodynamical critical field and, under the Kempf estimate, it is expected to be of the same order of magnitude of 
Casimir energy variation. Thus, as stated before, we assume the Kempf hypothesis and estimate the energy variation as totally due to 
Casimir effect. It is very important to stress that, as  will be shown in Secs. III and IV, even if the contribution of Casimir energy 
were of order of just a few over a 
thousand of the total energy at the transition, we might ascertain  whether it gravitates.

In the following sections the detection of small forces by using the best of 
current optical techniques will be considered. The use of high-$T_{c}$ 
superconductors in high sensitivity optical devices is a field yet to be 
investigated, in particular in macroscopic devices. Nevertheless,   
present superconductors can be deposited on quite large surface 
optical elements: YBCO is well deposited on aluminum 
(${\rm Al}_{2}{\rm O}_{3}$) substrates, which 
are the best substrates also for optics at low temperature. Indeed, a 
300 nm thick YBCO layer deposited on a 3-inches diameter, 5 mm thick 
${\rm Al}_{2}{\rm O}_{3}$ substrate produced by CERACO is presently under 
test in our laboratory. Notice that, even if the first test will be performed with YBCO for its robustness, the use of low upper 
critical fields supercondutors should be preferred, since they allow simpler magnetic modulation at equal values of thermodynamical 
critical field (see also Appendix B for definitions of thermodynamical and upper critical field). Furthermore, much larger thicknesses 
can be reached by using superconducting crystals. 

\section{Use of gravitational wave detectors}
\label{analisysII}

The force exerted by the gravitational field when the Casimir energy 
contained in the superconductor system is modulated should be compared with 
the up-to-date technology in the detection of small forces in macroscopic 
systems. Two main ways might be followed. The first way is to make use of 
the present most sensitive apparatuses in the detection of small forces, the 
gravitational wave detectors; the second is to go towards lower 
frequencies and use torsion pendulums. In the following we will consider 
first the use of gravitational wave detectors. 
The main reason to explore this way is the 
possibility of making use of a very well developed technology in force 
detection and seismic attenuation. Another not negligible reason is that 
money can be saved if a replica of many instruments and methods already 
available is avoided. In this case, it is necessary 
to recover an experimental method, discussed later, to apply a force on such 
detectors (only at a given frequency) without perturbing the gravitational 
wave measurement in the other frequencies of the spectrum.  
Our comparison can start with the present state of the art of gravitational 
wave detectors. Over the last decades, this field has known many impressive 
technical improvements and developments. The two most sensitive detectors 
of gravitational waves, LIGO and Virgo, have demonstrated the feasibility 
of all foreseen techniques, by reaching, and in some frequency regions 
superseding, the sensitivities expected for the first generation detectors 
\cite{virgo1,virgo2}. Moreover, many important techniques already compliant 
or extremely useful in the next generation detectors have been demonstrated 
worldwide, i.e. in LIGO \cite{ligo-enh}, Virgo \cite{virgosuper} or in the 
medium-scale detectors like GEO \cite{geo} or still in development like 
Kagra \cite{kagra}. In light of all this progress it is very reasonable to 
expect, for the second generation of such detectors, the so-called Advanced 
Detectors, presently under construction, to reach the design sensitivities 
in the next few years \cite{virgo-adv,ligo-adv}. 

In this case the frequency region of highest sensitivity ${\widetilde S}_F$ to 
the force lies in the range from 20 to 40 Hz; if a 
gravitational wave test mass of 42 Kg is 
considered, the value, in this region, is of order of 
${\widetilde S}_F \approx 10^{-13} N/\sqrt{Hz}$.  

Glancing at future detectors, the so-called third-generation detectors, 
like Einstein Telescope (ET), we see that they will benefit of low seismic sites, low temperature 
and suitably injected power for low-frequency detection. The expected 
sensitivity in the amplitude of the force will gain about two orders of 
magnitude, showing the region of best force sensitivity at frequencies 
slightly smaller than 10 Hz \cite{et}. 

\begin{figure}
\includegraphics[width=0.8\linewidth]{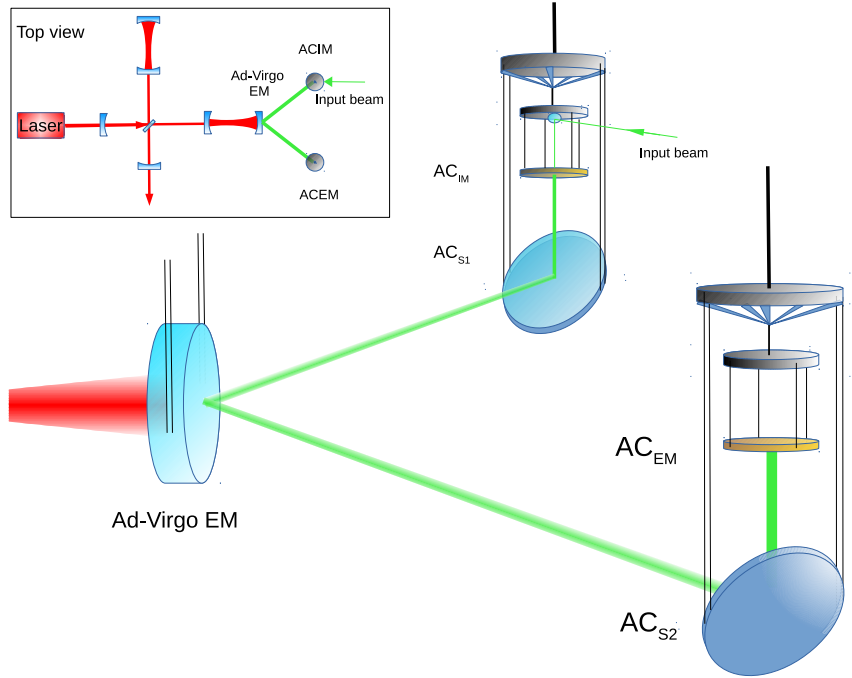}
\caption{Sketch of the optical link. The cavity acting as an optical spring is composed by the Archimedes Cavity input mirror, $\rm{AC_{IM}}$, by the steering mirror $\rm{AC_{S1}}$, by the back surface of the Gravitational wave detector test mass, AD-Virgo EM, by a second steering mirror $\rm{AC_{IM}}$ and closed by the Archimedes Cavity end mirror $\rm{AC_{IM}}$. Also a top view of the apparatus and the Gravitational wave detectors is sketched in the top-left box, not in scale, to show a complete view. The cavity is illuminated by a laser reflecting on the upper stage that suspends the input mirror, as shown in figure.}
\label{disegnoArchimedes}
\end{figure}

Two main conditions, in our opinion, constrain and define the 
use of gravitational wave detectors also for a 
measurement of the weight of vacuum. The first is that 
no modifications are allowed to the gravitational wave 
detector that in any case risk to 
reduce the gravitational wave sensitivity. 
In particular, no changes of the suspensions 
chain, of the payloads, of the actuators will be allowed: the system 
providing the force should be ``external" and sufficiently far from the 
gravitational wave test masses so as to avoid introducing spurious signals. 
The second is that the vacuum weight force is vertical (i.e. orthogonal 
to the earth's surface) while the gravitational detectors 
are designed to detect 
horizontal forces (i.e. almost parallel to the earth surface, with a small 
coupling factor with the vertical due to earth's curvature and mechanical 
imperfections).  

A possible way to face both points is to build an ad-hoc apparatus, 
lying several meters from the gravitational wave 
detectors test masses; let's call it the Archimedes system. In this system, a mass, over which the superconducting materialis deposited, is suspended and free to move vertically for frequencies above a few Hz. By applying the modulation technique discussed previuously, a force is exerted on the mass. To transport this force from the mass of the Archimedes system to the test mass of gravitational wave detector the ideal way would be to link them with a spring. It is not possible, for the reasons discussed above, to use a mechanical spring but,  as we shall see, it is possible to link the two masses via the radiation pressure, by realizing an optical cavity that, in a properly detuned configuration, acts as an optical spring \cite{corbitt}.

To show the behavior of an optical spring, let us consider a Fabry-Perot cavity with a suspended perfectly reflective end mirror,   and fixed highly reflective input mirror, and analyze it in the static approximation, valid for frequencies lower than cavity linewidth. Suppose that it is illuminated by a laser light with frequency
$\omega_{0}$ and power $I_0$.  Assuming the cavity to be close to
resonance, we list several quantities characterizing the state
of the cavity, i.e., its linewidth $\gamma$, finesse $F$, circulating
power W, and the phase shift $\Phi$ gained by the light as it
comes out from the cavity, in terms of more basic parameters:
\begin{equation}
\gamma = \frac{c T_{I}}{4L} ,
\end{equation}
\begin{equation}
\textsl{F} = \frac{2\pi}{T_I} ,
\end{equation}
\begin{equation}
W(I_0, \delta_{\gamma}) = \frac{4I_0}{T_I}\frac{1}{\left(1 + \delta_{\gamma}^2\right)} ,
\end{equation}
\begin{equation}
\Phi(\delta_{\gamma}) = -2{\rm tan}^{-1}( \delta_{\gamma}).
\end{equation}
Here L is the cavity length, $T_I$ the input-mirror power transmissivity. The
detuning parameter $\delta_\gamma$,
\begin{equation}
\delta_{\gamma} \equiv \frac{ \delta}{\gamma},
\end{equation}
is defined in terms of $\delta \equiv \omega_{\rm {res}}- \omega_0$, the difference between the cavity resonant frequency and laser frequency.
The ponderomotive force $F_p$, the radiation pressure, is given by
\begin{equation}
F_p = \frac{ 2W}{c}
\end{equation}
If the suspended mirror moves by an amount $\delta x$, since the cavity is not perfectly on resonance, the amount of light inside the cavity changes, and hence the radiation pressure on the mirror: a restoring force $F_r$ is produced  equal to $F_r = - K_{\rm{opt}}\, \delta x$, where
 $K_{\rm{opt}}$ is the optical spring constant, given by
\begin{equation}
K_{\rm{opt}} = \frac{2}{c}\frac{\partial W(I_0,\delta_\gamma)}{\partial \delta_\gamma} \frac{\partial \delta_\gamma}{\partial x} = 
-\frac{4 \omega_0 W }{\gamma L c}\frac{ \delta_\gamma }{\left(1 +  \delta_\gamma^2\right)}.
\end{equation}
With some algebra it can be written as
\begin{equation}\label{Theta}
K_{\rm{opt}}  = -\frac{4 \omega_0 I_0 \delta_\gamma}{ c^2} \left[\frac{2F}{\pi} \frac{1}{(1 + \delta_\gamma^2)}\right]^2
\end{equation}
The optical spring constant can be positive or negative, depending on the sign of the detuning $\delta_\gamma$. We choose a negative detuning so that the constant is positive. 
Remarkably, the optical spring constant, for sufficiently high finesse, can be quite high. For example, suppose to have a cavity with Finesse $F = 6\times 10^5$, input power $I_0$ = 0.16 mW, detuning $\delta_\gamma = -0.3$, laser frequency $\omega_{0} = 3\times 10^{14}$ Hz,  (corresponding to laser YAG wavelength of $1.064 \; \mu\rm{m}$), the optical spring constant is then equal to $K = 7.8\times 10^4$ N/m.
If the cavity is composed by two or more suspended mirrors a similar analysis applies and the light acts as a spring.\\
The other key feature of the optical spring is the low noise reintroduced: if we assume that the laser is shot noise limited the fluctuation power incident on the cavity is $\tilde{I_0} = \sqrt{2 \hbar  \omega_0 I_0}$ and  
 this induces a fluctuating noise force 
\begin{equation}
\tilde{F}_n = \frac{2}{c}\frac{\partial W(I_0,\delta_\gamma)}{\partial I_0} \tilde{I}_0 = \left(\frac{2F}{\pi}\frac{1}{(1 + \delta_\gamma^2)}\right)\frac{2\sqrt{2 \hbar \omega_0 I_0}}{c} = 6 \times 10^{-15} \, \frac{\rm{N}}{\sqrt{ \rm{Hz}}}.
\end{equation}

This small value of injected noise arises from the small amount of light that circulates in the cavity, even in presence of a high spring constant, a condition that can be reached by  using high Finesse cavities.
The actual apparatus is sketched in fig. \ref{disegnoArchimedes}: the cavity is composed by 5 optical elements. An input mirror coated with superconducting material, except for a small area to let the light pass. This  mirror has the surface parallel to ground. A 45 degrees reflective mirror lying below the input mirror, that sends the beam to the back surface of the gravitational wave detector test mass. The beam impinges upon the mirror at few degrees in the horizontal plane, hence it is reflected towards a second Archimedes apparatus that closes the cavity. 
The mirrors of the Archimedes apparatuses coated with superconductor, have masses m = $5$ Kg and are suspended to a seismic isolation sistem similar to Virgo ones. The $45$ degrees mirrors are suspended to the same attenuation system, but at an upper stage, to be independent of the mirrors coated with the superconductor: they act as merely deflection mirrors. They are quite heavy, of the same order of magnitude of the gravitational wave test mass.
The superconductor covers on the two faces of each coated mirror an area
$S = 0.23 \; m^{2}$, on each mirror, with a thickness of about 250 $\mu m$. The substrate is ${\rm Al}_{2}{\rm O}_{3}$, that is particular well suited to low temperature work. The area is similar to the present beam splitter of Virgo detector.

\begin{figure}
\includegraphics[width=0.8\linewidth]{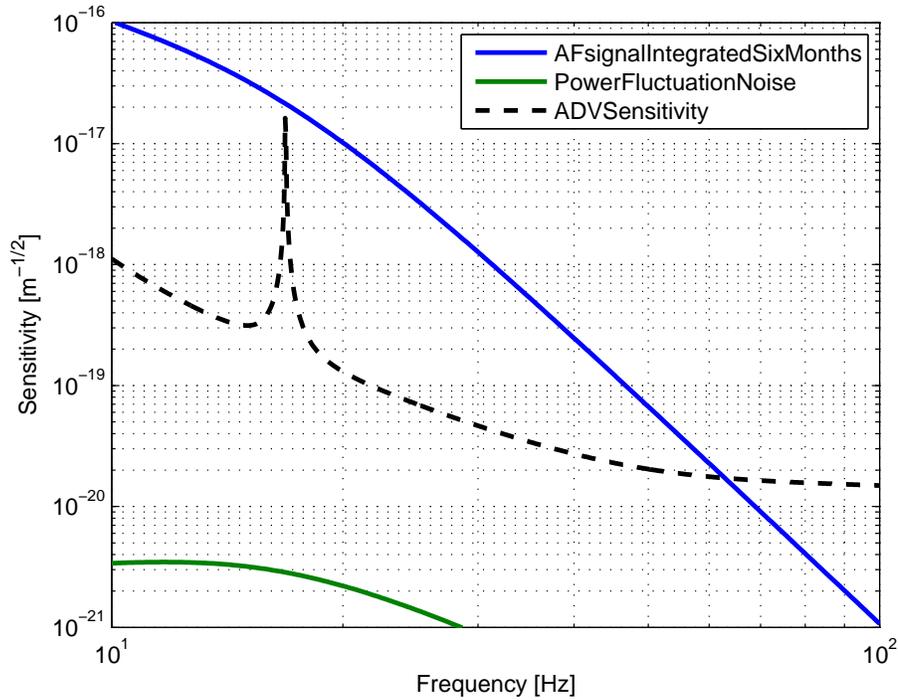}
\caption{Expected signal for the YBCO actuator described in the previous 
Section.}
\label{advsignal}
\end{figure}

The amplitude of force modulation $F_m$ can be evaluated as  
\begin{equation}
F_m \approx\, N \eta(a) \frac{E_{\rm(Cp)}}{c^2}g \approx \, {\cal N} \left[ -5 \times 10^{-3} \hbar 
\frac{c A}{ a^{5/2}} \sqrt{\Omega}\right]  \approx 10^{-15} N ,
\label{equazione_5)}
\end{equation}
where $\eta$ is the reduction factor with respect to the perfectly reflecting 
plates Casimir energy $E_{\rm(Cp)}$ \cite{kempf}, ${\cal N} = 1.6 \times 10^{5}$ is the total number of layers and $a = 1.17$ nm 
is the conducting layers separation in YBCO. 
To compare the effect of this force with the sensitivity of the gravitational wave detector we will compare the displacement induced in the gravitational test mass with respect to the displacement sensitivity. 
Note that, if the gravitational wave detector test mass is linked 
by an optical spring to other free masses, under the condition presently 
assumed of small distances, with respect to armlength, 
(and not considering the region of frequency 
around the optical spring resonance frequency) 
the displacement of the gravitational wave 
test mass induced by a gravitational wave will not change because all masses will accelerate at once.
Note that this statement also assumes that the masses of the
gravitational wave detector are free.  This is not strictly the case: the mass is linked by the arm-cavity optical spring to the rest of the masses of 
the gravitational wave interferometer. To reach a precise statement, and not an order of magnitude expectation,  a complete simulation of the Archimedes cavity coupled to the interferometer should be performed, which because of the complexity, is outside the 
aim of the present paper, and will be investigated in the near future.
At present, a complete simulation of the Archimedes cavity has been performed.
The system has been simulated by using the Optickle code \cite{matt}.
Under the assumptions and the parameters discussed above, the expected signal for an integration time of  6 months, a typical time-scale of a run, is given in Fig. \ref{advsignal}.
    
The signal is above the sensitivity by two orders of magnitude at low 
frequency, while it falls under the ADV sensitivity around 100 Hz. As 
expected, the noise due to power fluctuations is  negligible. Indeed, the power inside the cavity is about 60 Watt, to be compared 
with the 0.6 MW of light circulating in the gravitational wave arm cavity.
In conclusion, the use of the optical spring to transport the force from the 
actuator system to the gravitational wave mass makes it possible to locate 
the actuator several meters away from the gravitational wave detector mass, avoiding possible spurious interactions.
The suspension system of the Archimedes force apparatus can be 
a replica of the ones of gravitational wave detectors,  
the cryogenic system can benefit of the 
several experimental studies and realizations now making progress in 
the world \cite{kagra}. In this way, the optical system 
reduces to the optical spring actuator, which 
is relatively simple, being just a laser suitably locked on the cavity. 

\begin{figure}
\includegraphics[width=0.8\linewidth]{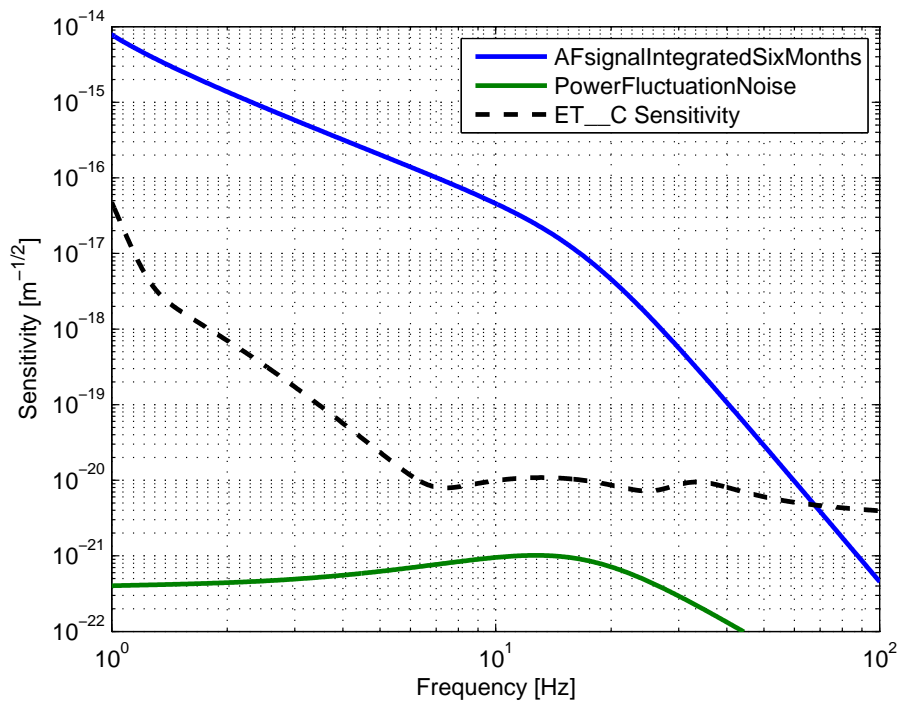}
\caption{Expected signal when using ET optimized for low frequency.}
\label{ETsignal}
\end{figure}

Note that, if the same system were applied to the next generation
of gravitational wave detectors, in particular Einstein Telescope low-frequency \cite{et,et2}, a remarkable improvement is expected. This is shown 
in Fig. \ref{ETsignal}. The cavity considered to perform the optical spring 
is similar to the previous one, with masses of 10 kg and a larger finesse 
= $1.5 \times 10^6$, that is not far from the current technological achievements. 
The input power is $P_{{\rm in}} = 1.6 \times 10^{-4}$ 
(not critical). The input power noise has been taken as shot-noise limit of 
the input power, equivalent to the noise-to-power ratio (RIN) of about 
$5 \times 10^{-8} \, 1/\sqrt{Hz}$: the power noise is more critical in this 
case but is negligible, remaining an order of magnitude lower than the 
sensitivity. The Figure shows that with an integration time of 6 months, 
the signal-to-noise ratio of about S/R = $10^4$ is reached. This means that 
a signal-to-noise ratio of 
1000 might be reached in a couple of days.  
    
With such high signal-to-noise ratios, measurements
with different materials, different layers separations up to tens of 
nanometers would then be possible, allowing a complete campaign of studies. 
This possibility clarifies also our working case on the Kempf 
hypotheis. According to that recipe, 
all the condensation energy, both at small and larger layer separation,  
results from Casimir energy. With this sensitivity, considering the 
accuracy of the gravitational wave detectors, even if the contribution of 
Casimir energy were only a few parts over a thousand, we might test 
whether it gravitates.
     
\section{Use of balances}
\label{analisysIIa}
The use of balances might be favored by the possibility to go 
towards low frequencies. Indeed, the modulation of superconducting phase 
transitions in macroscopic bodies, is expected to be easier at lower 
frequencies. Furthermore, we will consider here the possibility 
of performing force modulation also by temperature modulation. We evaluate the thermal noise at the temperature working point of 100 K, 
near the YBCO transition temperature. The main experimental point that has to be faced in going towards low frequencies is that a proper seismic 
attenuation system for balances does not yet exist. 

\begin{figure}
\includegraphics[width=0.8\linewidth]{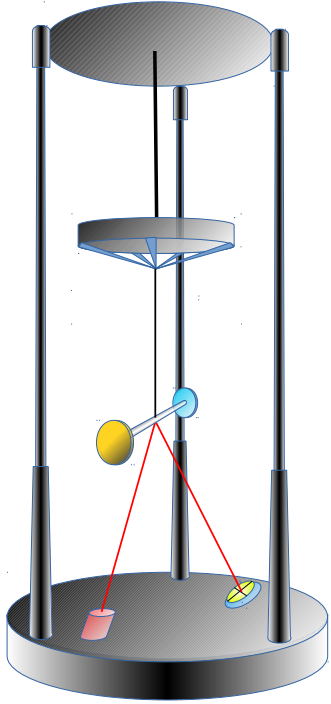}
\caption{Sketch of the balance with the optical lever detection read-out. The seismic attenuation chain is composed by the three-legs inverted pendulum and by the blades-attenuation element. The read-out is composed by a laser beam that reflects on the bottom of the bar and impinges on a quadrant photodiode.}
\label{disegnoBilancia}
\end{figure}

A possible way to reduce seismic noise at frequencies lower than 0.1 Hz is to 
hang the balance to a cascade formed by an inverted pendulum followed by a 
blades' isolation stage, like shown in Fig. \ref{disegnoBilancia}. The inverted pendulum is efficient in the two 
horizontal translational degrees of freedom and the rotation around the 
vertical axis, while the blades' stage is efficient in the vertical degree 
of freedom and in the rotations \cite{harms,virgosuper}. The Virgo inverted 
pendulum has already demonstrated to have a resonant frequency of 0.03 mHz 
and work is ongoing to further reduce it to the value of 0.01 Hz. Also the 
blades' stage resonance can be tuned, by careful tuning of magnetic 
antispring stiffness, to similar values.

The control of this top stage can be done either at very low frequency, 
with unity gain of the feedback lower than the resonance or in high 
bandwidth, with unity gain of about 1 Hz.
Here we assume to close the loop in high gain and reach, at the suspension 
point of the balance, the electronic noise floor of the accelerometers 
$a_s \approx 4 \times 10^{-10} m^{2}/s \sqrt(Hz)$, 
corresponding to the displacement 
noise of  $1nm/\sqrt(hz)$ at 0.1 Hz, and flat for frequency 
less than 0.1 Hz \cite{harms}.   
To calculate the expected signal and noises at the balance, we have 
considered a balance having arms of length L = 0.1 m, a plate at each 
arm's end of mass $M = 0.4$ kg, total mass $M_b = 1.25$ kg, moment of inertia 
$I = 0.01 \rm{kg \, m^{2}}$, resonance frequency 
$F_{\rm{res}} = \omega_b/2\pi = 5$ mHz, with mechanical internal loss angle $\phi = 10^{-6}$.
The resonance value is higher than typical torsion pendulum (horizontal) 
ones already existing \cite{trento} and takes into account the feasibility 
of a real vertical balance: in particular, the resonance of 5 mHz corresponds 
to careful setting of the bending point distance from the balance center of 
mass of about $h_b \approx 1 \mu m$.  (The bending point is the physical point around which the balance rotates. Its position depends upon the point were the wire is fixed on the balance, the mass of the balance and the wire section and Young modulus. The distance $h_b$ of the bending point from the balance center of mass determines the balance's resonance frequency $\omega_b$, with the relation $\omega_b^2 = \frac{M_b g\,h_b}{I}$. This distance can be tuned both mechanically, by regulating ballasts' position, and in feed-back, with the help of external forces.)

\begin{figure}
\includegraphics[width=0.8\linewidth]{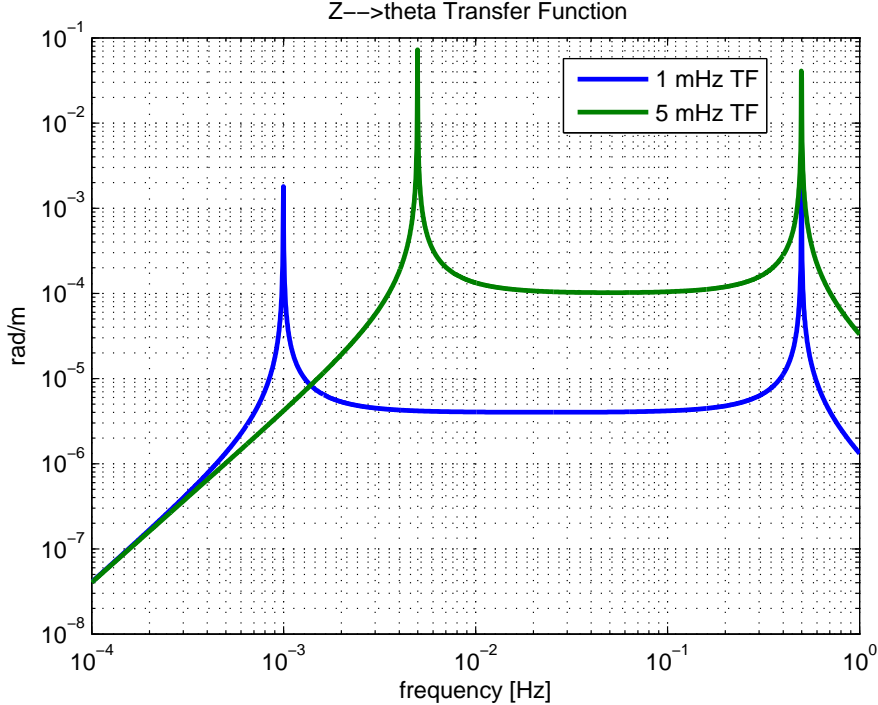}
\caption{Transfer function from displacement suspension point to balance's 
tilt for 1mHz and 5 mHz resonance frequency. The coupling depends on the 
resonant frequency.}
\label{TFzt}
\end{figure}

The material to be used for the suspension fiber (and for the balance itself) 
cannot be fused silica, which is the material of choice for the test masses 
of all first-generation gravitational wave detectors, 
because it has a high dissipation at 
low temperatures \cite{glass-thermal,rowan}. Sapphire has already been 
proposed as alternative material also for the suspension fiber, 
and here we assume it is the final material 
\cite{sapphire-thermal}. The wire length considered in our simulation is 
of 1 m and the diameter d = 50 $\mu m$. 
The end plates have radius R=0.15 m, made by a sapphire substrate and one 
is coated with 250 $\mu m$ of YBCO on both faces: the force modulation on 
the plate is  $F_a = 4 \times 10^{-16}$ N.
As expected, simulations show that the most critical noise is the seismic 
noise injected through the coupling of transversal motion of suspension point to tilt of the balance. The simulated 
transfer function is shown in Fig. \ref{TFzt} for the case of 5 mHz and 
for a very optimistic case, similar to torsion pendulum value, of 1 mHz
to show that this parameter is critical for reaching a significant 
attenuation.

\begin{figure}
\includegraphics[width=0.9\linewidth]{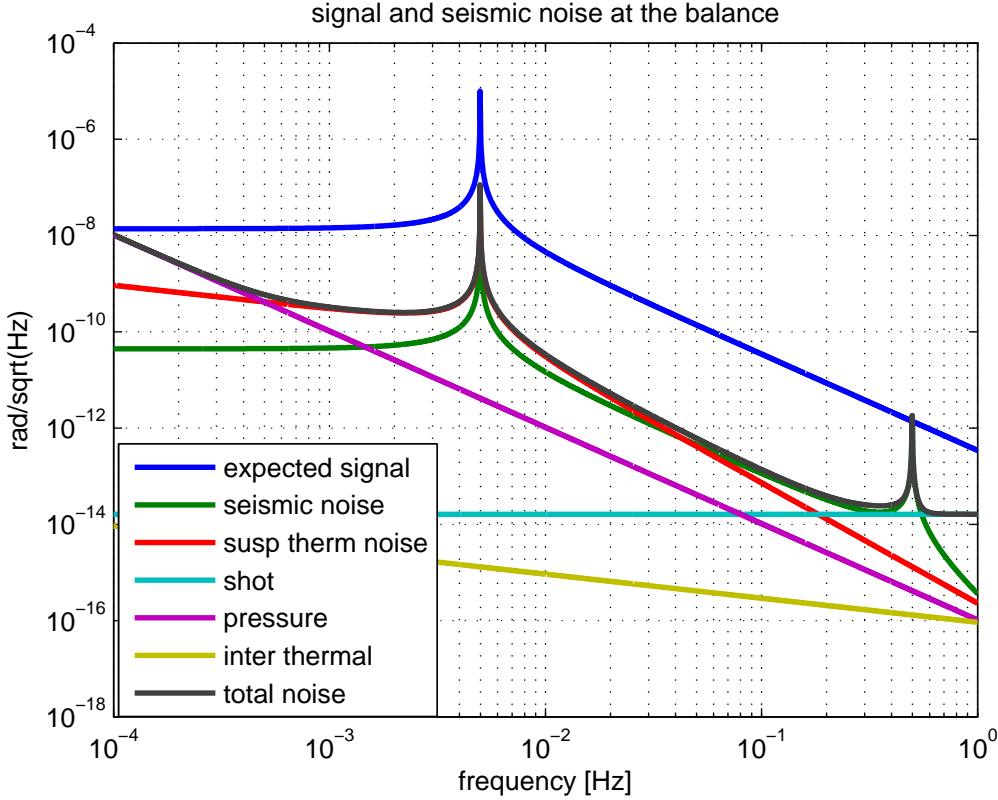}
\caption{Expected signal and noises for the balance. In the
region of frequencies $5 < f < 100$ mHz the signal is about two orders 
of magnitude above the noise.}
\label{sensitivityAndNoises}
\end{figure}

The tilt signal can in principle be read off in various ways. A 
high-sensitivity possibility is to use a second balance and read the ends' 
differential displacements with a Michelson interferometer having a 
Fabry-Perot cavity at the ends of the balances' arms. For an interferometer 
having arm cavity finesse Fb = 100,  input power Pb = 0.01 W, the sensitivity 
is reported in Fig. \ref{sensitivityAndNoises} where the radiation pressure 
noise and shot noise are plotted. The signal (blue curve) is obtained 
by integrating for 6 months and is approximately two orders of magnitude 
larger than the total noise (black curve). 

Under the assumption on seismic noise reduction the sensitivity is limited 
at low frequency by suspension thermal noise and by seismic noise for 
frequencies larger than 30 mHz.  
The radiation pressure noise and shot noise curves ensure that 
fundamental noises will not make it impossible 
to perform the measurement of the vacuum-gravity force. Nevertheless, the 
noise is so lower with respect to other noises that other tilt 
detection methods, even if more noisy, can be exploited if simpler. 
As an example, optical lever systems or capacitors used in torsion pendulums 
have already shown remarkable sensitivities; they are not yet fully 
compatible with our needs, but surely deserve careful study and attention 
\cite{trento,difiore}. The use of such detection system is also sketched in Fig. \ref{disegnoBilancia}:  a low power laser beam is sent to the balance and reflected towards a quadrant photodiode; a tilt of the balance displaces the impinging point of the beam on the photodiode and a signal is hence generated.

\begin{figure}
\includegraphics[width=0.9\linewidth]{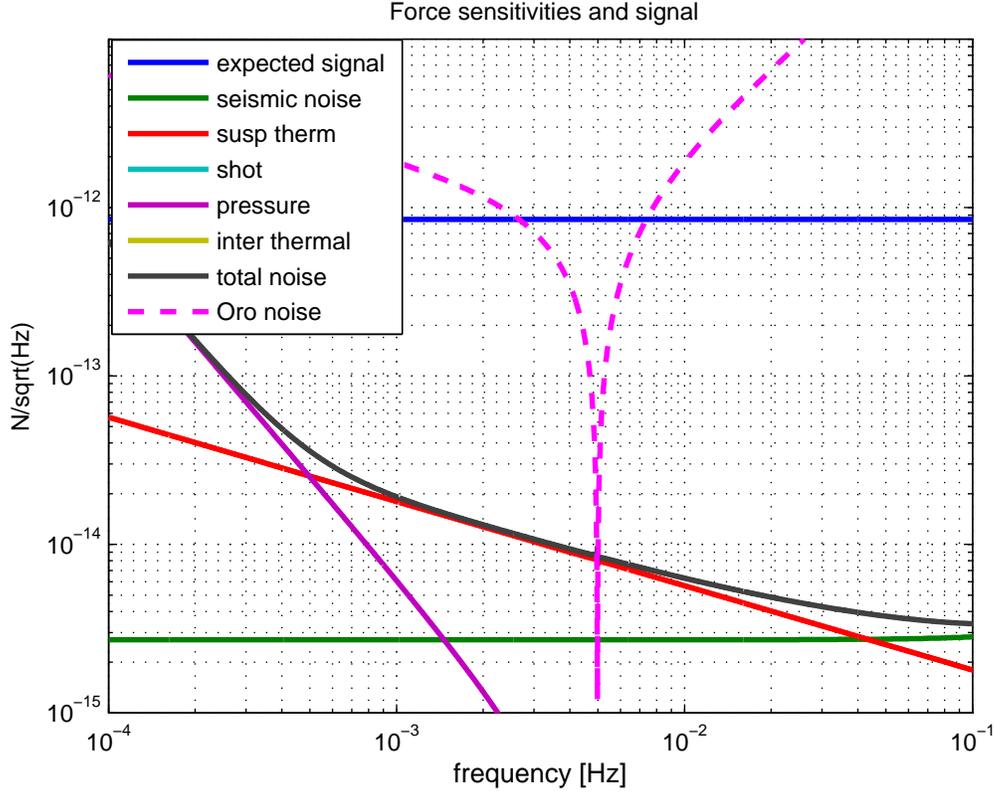}
\caption{Force signal and noises. The dashed line describes the noise of 
an optical lever detection system.}
\label{sensitivityForce}
\end{figure}

The corresponding signal and noises in $N/\sqrt{Hz}$ are plotted in 
Fig. \ref{sensitivityForce}. The coupling of suspension point 
acceleration $a_s$ can be interpreted as producing a moment of inertia 
$M_s = M_b\cdotp a_s \cdotp h_b$, equivalent to the noise force 
$M_b \cdotp a_s \cdotp h_b/L_b$, that again shows how the setting of the 
bending point is critical. The plot reports (dashed line) also the 
read-out noise of an optical lever demonstrated in Ref. \cite{difiore}:  
such a system makes it mandatory to perform the measurement 
in the neighborhood of the resonance (at the price of slightly reducing the sensitivty), but leads to a remarkable simplification 
of the detection method.

\section{Conclusions}
\label{last}

We have shown that it is by now possible to begin the experimental path 
to check against observation whether virtual photons 
do gravitate and to verify the Archimedes force of vacuum. 
Various experimental techniques must be investigated and refined, i.e., 
deposition of thick layers of high-$T_{c}$ superconductors in 
optical substrates, application of optical springs to connect different 
apparatuses, improvements in low-frequency seismic isolation. If these 
improvements, not far from the present technological achievements, will be 
successful, a first answer will be given to one of the deepest and 
longlasting problems of fundamental physics.

\begin{acknowledgments}
L. Di Fiore and G. Esposito are grateful to Dipartimento di Fisica of Federico II University for hospitality and support. 
Previous collaboration with G. Bimonte on several theoretical aspects has been of great importance. We are grateful to 
A. Coltorti, M. Cerdonio, F. Fidecaro, F. Frasconi, E. Majorana, P. Rapagnani, F. Ricci for useful discussions.
\end{acknowledgments}

\appendix
\section{Dielectric conductivity in Type-I superconductors}

In the case of a BCS conductor at a
temperature $T<T_c$, the expression of $\sigma'_s(\omega)$ can be written as
\begin{equation}
\sigma'_s(\omega)=\kappa \,\delta(\omega)+\hat{\sigma}'_s(\omega)
\;.\label{bcssum}
\end{equation}  
For $\omega > 0$, $\hat{\sigma}'_s(\omega)$
reads as \cite{berl} 
\begin{equation} 
\hat{\sigma}'_s(\omega)=\frac{\hbar \,
n\, e^2}{2 m \omega
\tau_n}\left[\int_{\Delta}^{\infty}dE\,J_T+\theta(\hbar \omega-2
\Delta)\int_{\Delta -\hbar \omega}^{-\Delta}dE\,\,J_D
\right]\;,\label{bcscon}
\end{equation}
 where
\begin{eqnarray}
J_T &:=& \,g(\omega,\tau_n,E)\,\left[\tanh \frac{E+\hbar \omega}{2 K T}-
\tanh \frac{E}{2KT}\right], \\
J_D &:=& -g(\omega,\tau_n,E)\;\tanh \left(\frac{E}{2K T}\right)\;,
\end{eqnarray}
with $K$ the Boltzmann constant. Defining \begin{equation} P_1:=\sqrt{(E+\hbar
\omega)^2-\Delta^2}\;,\;\;\;\; P_2:=\sqrt{E^2-\Delta^2}\;, \end{equation} the
function $g(\omega,\tau_n,E)$ is
\begin{eqnarray}
g &:=& \left[1+\frac{E(E+\hbar \omega)+\Delta^2}{P_1 P_2}\right]\frac{1}{(P_1-P_2)^2+(\hbar/\tau_n)^2}
\nonumber\\
&-&\left[1-\frac{E(E+\hbar \omega)+\Delta^2}{P_1 P_2} \right]\frac{1}{(P_1-P_2)^2+(\hbar/\tau_n)^2}
\;.\nonumber
\end{eqnarray}
The  coefficient $\kappa$ of the Dirac delta in Eq.
(\ref{bcssum}) is determined so as to satisfy the sum rule
 \begin{equation} 
\int_0^{\infty} d \omega
\;\sigma'(\omega)= \frac{\pi n e^2}{2m}\;,\label{srule}
\end{equation}
where
$n=n_s+n_n$ is the total electron density and can be computed exactly according to
\cite{berl}
\begin{widetext}
\begin{equation}  
\kappa=\frac{\pi n e^2}{m}
\;  \left[\frac{\pi \tau_n \Delta}{\hbar} \tanh \frac{\Delta}{2 K T}
-4 \Delta^2 \int_{\Delta}^{\infty} dE\;\frac{\tanh(E/2 K T)}
{\sqrt{E^2-\Delta^2}[4(E^2-\Delta^2)+(\hbar/\tau_n)^2]}\right]\;.\label{delta}
\end{equation}
\end{widetext}

\section{Internal energy variation in type-II superconductors' transitions}

Measuring the variation of weight of the superconductor when it undergoes a transition means to measure the variation of its internal 
energy among the two states, normal and superconducting. The internal energy difference of the system in different states can be evaluated 
by means of the thermodynamical potentials  $H'$ (enthalphy), $G$ (Gibbs free energy) and $S$ (entropy).

For a magnetic material the differential of the internal energy $dU$ may be written in terms of the temperature $T$, the applied 
magnetic field $\mathbf{B}$, and the magnetization $\mathbf{M}$ of the material as
\begin{equation}
dU = TdS + \mathbf{B} \cdotp \mathbf{dM} .
\end{equation}\label{du}
The enthalpy is defined as
\begin{equation}
H' \equiv U - \mathbf{B} \cdotp \mathbf{M},
\end{equation}\label{h}
and finally the Gibbs free energy as
\begin{equation}
G \equiv H' - TS,
\label{g}
\end{equation}
with differential form
\begin{equation}
dG = -SdT - \mathbf{M} \cdotp \mathbf{dB} . 
\label{dg}
\end{equation}

In type-II superconductors, for applied field $\mathbf{B}(T)$, where $T$ is the temperature, they are defined as the lower critical field, 
such that if $B \lneq B_{c1}(T)$ the field does not penetrate in the sample, and the upper critical field $B_{c2}(T)$ such that if  
$B \geq B_{c2}(T)$, the superconductivity is destroyed. In analogy with type-I superconductors, a thermodynamical critical field  
${\mathbf B}_{c}(T)$ is defined such that the difference of Gibbs free energies, at given temperature, among the superconducting and normal 
states at zero applied field is
\begin{equation}
G_s(T,0) - G_n(T,0) = \frac{(B_c(T))^2}{2\mu_0}. 
\label{gsn} 
\end{equation} 
Following the Ginzburg-Landau theory the lower critical field $B_{c1}$ and upper critical field $B_{c2}$ are linked to the 
thermodynamical critical field $B_c$ by the dimensionless parameter $k = \frac{\lambda(T)}{\xi(T)}$: 
\begin{equation}
B_{c1} = \frac{B_c\,\log{k}}{\sqrt{2}k},
\end{equation}
and 
\begin{equation}
B_{c2} = B_c\sqrt{2}k.
\end{equation}
The temperature dependence of the critical field $B_c(T)$ is well approximated by 
\begin{equation}
B_c(T) = B_c(0)\left[ 1 - \left(\frac{T}{T_c}\right)^2 \right ],
\end{equation}
and similarly for $B_{c1}(T)$ and $B_{c2}(T)$.

In high-$T_{c}$ superconductors like YBCO, $k$ is of order of 100. The entropy of the superconductor in applied field, 
to take heuristically into account the magnetization of the superconductor and fit experimental data \cite{bonjour} is approximated as
\begin{equation}
S_s(T,B) = S_n(T)+ \chi '(T) \frac{(B_{c2}(T) - B)}{\mu_0} \frac{dB_{c2}}{dT},
\end{equation}
where $\chi ' (T) = \mu_0\frac {\partial M}{\partial B}$ is called the differential susceptibility. It takes into account the 
anisotropy of type-II superconductors and maintains the entropy at vanishing field independent of the anisotropic value of $B_{c2}$. 
This expression for the entropy shows that in type-II superconductors, unlike the type-I case, the transition obtained by applying an 
external field at fixed temperature $T \leq T_c$, is of second order, with no latent heat. The transition in vanishing field, 
for $T = T_c$ is of second order, since $B_{c2}(T_c) = 0$. Thus, both modulation techniques proposed are in absence of latent heat.
Let us observe that, if the applied field $B \lneq B_{c1}(T)$ and the transition is  obtained by increasing the temperature $T$, the 
superconductor behaves like a type-I superconductor (the field does not penetrate and there is an entropy variation).  
The transition is thus of first order, with latent heat $Lh$ equal to
\begin{equation}
Lh = T_c(B)\, \left(S_n - S_s\right) =  
\frac{B_{c1}^2}{2\mu_0}\left(\frac{T_c(B)}{T_c}\right)^2 \times 
\left[ 1 - \left(\frac{T_c(B)}{T_c}\right)^2 \right].
\end{equation} 

As stated in Sec. II, the transitions considered in the present paper are two: the first is a transition by temperature variation in 
vanishing field. The second is a field variation up to $B_{c2}(T)$ at constant $T$. Both cases have no latent heat.
To evaluate the internal energy variation $\Delta U$, in the first case, considering vanishing field and noticing that 
$G_s(T_c) = G_n(T_c)$, we can write
\begin{equation}
S_s(T) - S_n(T) = -\frac{dG_s(T,0)}{dT} + \frac{dG_n(T,0)}{dT} = \frac{1}{2\mu_0}\frac{d\left[B_c(T)\right]^{2}}{dT} 
= \frac{B_c(T)}{\mu_0}\frac{d B(T)}{dT}.   
\end{equation}

In vanishing field the variation of internal energy $\Delta U$ is equal to the variation of enthalpy. 
By using (\ref{g}) this variation can be written as
\begin{eqnarray}
\Delta U = H'_n(T_c) - H'_s(T) = H'_n(T,0) + \int_T^{T_c}C_ndT - H'_s(T,0) 
\nonumber \\  
=  \int_T^{T_c}C_ndT + G_n(T) + TS_n(T) - G(T,0) - TS_s(T,0)  
\nonumber \\  
=  \int_T^{T_c}C_ndT + \frac{B_c(T)^2}{2\mu_0} - T\frac{B_c(T)}{\mu_0}\frac{dB_c(T)}{dT}. &
\end{eqnarray}  
Considering that $B_c(T) = B_0\left[1 - (T/T_c)^2\right]$, the variation of energy can be written as the sum of three terms: 
the internal energy variation of normal state, the contribution of the Gibbs free energy and the contribution of entropy. The third term, 
for temperatures near $T_{c}$, gives the biggest contribution, and one has
\begin{equation}
\Delta U = \int_T^{T_c}C_ndT + \frac{B_0^2}{2\mu_0}\left[1 - (T/T_c)^2\right]^2 
+ 2 \left(\frac{T}{T_c}\right)^2 \left[1 - (T/T_c)^2\right].
\label{dufinal}  
\end{equation}
This equation shows that the variation of internal energy is proportional to, and roughly of the same order of magnitude as, 
the energy of the thermodynamical critical field. 

The second transition is provided by keeping the temperature $T$ fixed and by varying the applied field from zero to the critical field 
$B_{c2}(T)$. Notice that, being the energy scale given by the thermodynamical critical field, use of low-$k$ materials 
should be preferred, to maintain the upper critical field manageable.    
Notice that in this transition the magnetization of the sample is zero both at the start and at the end of the transition, 
hence it is expected that the variation of internal energy due to the superconductive contribution is equal to the previous case. 
The normal-state contribution, on the contrary, is zero because there is no temperature variation.
The above expectations can be verified by noticing that $U = G + \mathbf{B} \cdotp \mathbf{M} + TS $. The differential reads
\begin{eqnarray}
dU &=& TdS + \mathbf{B} \cdotp \mathbf{dM} = TdS + \mathbf{B} \cdotp \mathbf{dM} + \mathbf{M} \cdotp \mathbf{dB} 
- \mathbf{M} \cdotp \mathbf{dB} + SdT - SdT 
\nonumber \\
&=& d(TS) + d(\mathbf{M} \cdotp \mathbf{B}) + dG.
\end{eqnarray}
By integration among the two final states we obtain
\begin{equation}
\Delta U = T\left(S_n(T) - S_s(T)\right) + G_n(T) - G_s(T,0) ,
\end{equation}
which gives the same result of (\ref{dufinal}) without the contribution of the normal state.

\end{document}